\documentstyle[12pt]{article}
\begin{document}
\begin{titlepage}
\begin{flushright}
IC/2001/39\\
hep-th/0105249
\end{flushright}
\vspace{10 mm}

\begin{center}
{\Large The Cardy-Verlinde Formula and\\ 
Charged Topological AdS Black Holes}

\vspace{5mm}

\end{center}

\vspace{5 mm}

\begin{center}
{\large Donam Youm\footnote{E-mail: youmd@ictp.trieste.it}}

\vspace{3mm}

ICTP, Strada Costiera 11, 34014 Trieste, Italy

\end{center}

\vspace{1cm}

\begin{center}
{\large Abstract}
\end{center}

\noindent

We consider the brane universe in the bulk background of the charged 
topological AdS black holes.  The evolution of the brane universe is 
described by the Friedmann equations for a flat or an open FRW-universe 
containing radiation and stiff matter.  We find that the temperature and 
entropy of the dual CFT are simply expressed in terms of the Hubble 
parameter and its time derivative, and the Friedmann equations coincide 
with thermodynamic formulas of the dual CFT at the moment when the brane 
crosses the black hole horizon.  We obtain the generalized Cardy-Verlinde 
formula for the CFT with an R-charge, for any values of the curvature 
parameter $k$ in the Friedmann equations. 

\vspace{1cm}
\begin{flushleft}
May, 2001
\end{flushleft}
\end{titlepage}
\newpage

Recently, there has been growing interest in holographic bounds in cosmology 
\cite{el,ven,br,ram,kl,bou}, after the initial work by Fischler and Susskind 
\cite{fs}.  The holographic bound in its original form is found to be 
violated by the closed Friedmann-Robertson-Walker (FRW) universe \cite{fs} 
and the later works attempted to circumvent such a problem through various 
modifications.  Later, Verlinde \cite{ver1,ver2} proposed an ingenious 
holographic bound for a radiation dominated closed FRW universe which remains 
valid throughout the cosmological evolution.  The Verlinde's holographic 
bound unifies the Bekenstein bound \cite{bek} for an weakly self-gravitating 
universe and the Hubble bound \cite{el,ven,kl,bou} for a strongly 
self-gravitating universe in an elegant way.  (Cf. It is shown in Refs. 
\cite{bv,bfv} that the Verlinde's entropy bound is equivalent to causal 
entropy bound, which is a covariant formulation of the Hubble entropy bound.) 
Furthermore, it is shown \cite{ver1} that the Cardy formula \cite{car} for a 
two-dimensional conformal field theory (CFT) can be generalized to an 
arbitrary spacetime dimensions and such generalized formula, called the 
Cardy-Verlinde formula, coincides with the Friedmann equation at the moment 
when the proposed cosmological bound is saturated.  Such results were 
generalized to brane universes in various bulk backgrounds 
\cite{was,kps1,cai,bm,bim,kps2,youm,cai2,noo,caz}.  
The quantum effects to the Cardy-Verlinde formula were studied in Refs. 
\cite{odi1,odi2,odi3}.  In our previous work \cite{youm1}, we showed that the 
results of Refs. \cite{ver1,ver2} continue to hold even for an open and a flat 
radiation dominated FRW universes induced on a brane moving in the bulk 
background of the topological AdS-Schwarzschild black holes.  In this note, 
we study a generalization to the brane universe in the bulk background of 
the charged topological AdS black holes.  (The results in this paper can be 
straightforwardly generalized to the multi-charged topological AdS black 
holes in string theories.)  The holographic dual theory corresponds to the 
CFT with an R-charge.  We show that the temperature and entropy of the CFT 
are still simply expressed in terms of the Hubble parameter and its time 
derivative when the brane crosses the black hole horizon.  We find that the 
Friedmann equations coincide with thermodynamic formulas of the CFT at the 
moment when the brane crosses the black hole horizon.  We generalize the 
Cardy-Verlinde formula to include the contribution from the R-charge of the 
CFT, for any values of the curvature parameter $k$ in the Friedmann 
equations.  

The charged topological AdS black hole solution in $(n+2)$-dimensions has 
the following form \cite{man,zca,blp,man1,cjs,cas,cejm}:
\begin{eqnarray}
ds^2_{n+2}&=&-h(a)dt^2+{1\over{h(a)}}da^2+a^2\gamma_{ij}(x)dx^idx^j,
\cr
h(a)&=&k-{{w_{n+1}M}\over a^{n-1}}+{{nw^2_{n+1}Q^2}\over{8(n-1)a^{2n-2}}}
+{a^2\over L^2},
\cr
\phi&=&{n\over{4n-4}}{{w_{n+1}Q}\over a^{n-1}},\ \ \ \ \ \ 
\omega_{n+1}={{16\pi G_{n+2}}\over{n{\rm Vol}(M^n)}},
\label{tcadsbh}
\end{eqnarray}
where $\gamma_{ij}$ is the horizon metric for a constant curvature manifold 
$M^n$ with the volume ${\rm Vol}(M^n)=\int d^nx\sqrt{\gamma}$, $G_{n+2}$ is 
the $(n+2)$-dimensional Newton's constant, $M$ is the ADM mass of the black 
hole, $Q$ is the electric charge, $\phi$ is the electrostatic potential 
(difference between the horizon and infinity), and $L$ is the curvature 
radius of the background AdS spacetime.  The horizon geometry of the black 
hole is elliptic, flat and hyperbolic for $k=1,0,-1$, respectively.  The 
Bekenstein-Hawking entropy $S$, the chemical potential $\phi_H$ and the 
Hawking temperature ${\cal T}$ of the black hole are
\begin{eqnarray}
S&=&{{a^n_H{\rm Vol}(M^n)}\over{4G_{n+2}}},\ \ \ \ \ \ \ \ \ \ \ \ 
\phi_H={n\over{4n-4}}{{w_{n+1}Q}\over a^{n-1}_H},
\cr
{\cal T}&=&{{h^{\prime}(a_H)}\over{4\pi}}={{(n+1)a^2_H+(n-1)kL^2}\over
{4\pi L^2a_H}}-{{nw^2_{n+1}Q^2}\over{32\pi a^{2n-1}_H}},
\label{thermqnts}
\end{eqnarray}
where $a_H$ is the horizon, defined as the largest root of $h(a)=0$, and the 
prime denotes the derivative w.r.t. $a$.  

We obtain the effective Friedmann equations describing the evolution of the 
brane universe in the above bulk background (\ref{tcadsbh}).  In terms of a 
new time coordinate $\tau$, satisfying
\begin{equation}
{1\over{h(a)}}\left({{da}\over{d\tau}}\right)^2-h(a)\left({{dt}\over
{d\tau}}\right)^2=-1,
\label{costim}
\end{equation}
the induced metric of the $n$-brane takes the standard Robertson-Walker form
\begin{equation}
ds^2_{n+1}=-d\tau^2+a^2(\tau)\gamma_{ij}dx^idx^j,
\label{rwmet}
\end{equation}
with the cosmic scale factor $a$.  Making use of Eq. (\ref{costim}) and the 
following equation of motion \cite{ver2} for the brane action:
\begin{equation}
{{dt}\over{d\tau}}={{\sigma a}\over{h(a)}},
\label{eqbranact}
\end{equation}
where the parameter $\sigma$ is related to the brane tension, we obtain the 
following Friedmann equation:
\begin{equation}
H^2={{\omega_{n+1}M}\over a^{n+1}}-{{nw^2_{n+1}Q^2}\over{8(n-1)a^{2n}}}
-{k\over a^2}, 
\label{frdeq1}
\end{equation}
after setting $\sigma=1/L$.  Here, $H\equiv \dot{a}/a$ is the Hubble 
parameter, where the overdot denotes the derivative w.r.t. $\tau$.  Taking 
the $\tau$-derivative of Eq. (\ref{frdeq1}), we obtain the second Friedmann 
equation
\begin{equation}
\dot{H}=-{{n+1}\over 2}{{\omega_{n+1}M}\over{a^{n+1}}}+{{n^2w^2_{n+1}Q^2}
\over{8(n-1)a^{2n}}}+{k\over a^2}.
\label{frdeq2}
\end{equation}
So, the motion in the bulk background (\ref{tcadsbh}) induces radiation 
matter $\sim M/a^{n+1}$ and stiff matter $\sim Q^2/a^{2n}$ \cite{bm} on 
the brane universe.  

Making use of the fact that the metric for the boundary CFT can be 
determined only up to a conformal factor \cite{gkp,wit2}, we rescale 
the boundary metric for the CFT to be of the following form:
\begin{equation}
ds^2_{CFT}=\lim_{a\to\infty}\left[{L^2\over a^2}ds^2_{n+2}\right]
=-dt^2+L^2\gamma_{ij}dx^idx^j.
\label{bndrmet}
\end{equation}
It is argued in Ref. \cite{wit} that thermodynamic quantities of the CFT at 
high temperature can be identified with those of the bulk AdS black hole.  
Note, the CFT time is rescaled by the factor $L/a$ w.r.t. the AdS time.  
So, the energy $E$, the chemical potential $\Phi_H$ and the temperature 
$T$ of the CFT are rescaled by the same factor $L/a$:
\begin{eqnarray}
E&=&M{L\over a},\ \ \ \ \ \ \ \ \ \ 
\Phi_H=\phi_H{L\over a}={n\over{4n-4}}{{w_{n+1}QL}\over{a^{n-1}_Ha}}, 
\cr
T&=&{\cal T}{L\over a}={1\over{4\pi a}}\left[(n+1){a_H\over L}+(n-1){{kL}
\over a_H}-{{nw^2_{n+1}Q^2L}\over{8a^{2n-1}_H}}\right],
\label{entempads}
\end{eqnarray}
whereas the entropy $S$ of the CFT is given by the Bekenstein-Hawking entropy 
(\ref{thermqnts}) of the black hole without re-scaling.  In terms of the 
energy density $\rho=E/V$, the pressure $p=\rho/n$, the charge density 
$\rho_Q=Q/V$ and the electrostatic potential $\Phi=\phi{a\over L}$ of the 
CFT within the volume $V=a^n{\rm Vol}(M^n)$, the Friedmann equations 
(\ref{frdeq1},\ref{frdeq2}) take the following forms:
\begin{equation}
H^2={{16\pi G}\over{n(n-1)}}\left(\rho-{1\over 2}\Phi\rho_Q\right)
-{k\over a^2},
\label{frd1}
\end{equation}
\begin{equation}
\dot{H}=-{{8\pi G}\over{n-1}}\left(\rho+p-\Phi\rho_Q\right)+{k\over a^2},
\label{frd2}
\end{equation}
where $G=(n-1)G_{n+2}/L$ is the Newton's constant on the brane.  So, the 
cosmological evolution is determined by the energy density $\rho$ and the 
pressure $p$ of the CFT plus those of the electric potential energy due to 
the R-charge.  From these Friedmann equations, we obtain the following 
energy conservation equation:
\begin{equation}
d(\rho-\textstyle{1\over 2}\Phi\rho_Q)/d\tau+n(\rho+p-\Phi\rho_Q)=0.
\label{enrgcsv}
\end{equation}

The Friedmann equations (\ref{frd1},\ref{frd2}) can be respectively put 
into the following forms resembling thermodynamic formulas of the CFT:
\begin{equation}
S_H={{2\pi}\over n}a\sqrt{E_{BH}[2(E-\textstyle{1\over 2}\Phi Q)-kE_{BH}]},
\label{fredcft1}
\end{equation}
\begin{equation}
kE_{BH}=n(E+pV-\Phi Q-T_HS_H),
\label{fredcft2}
\end{equation}
in terms of the Hubble entropy $S_H$ and the Bekenstein-Hawking energy 
$E_{BH}$, where
\begin{equation}
S_H\equiv (n-1){{HV}\over{4G}}, \ \ \ \ \ \ \ 
E_{BH}\equiv n(n-1){V\over{8\pi Ga^2}}, \ \ \ \ \ \ \ 
T_H\equiv -{\dot{H}\over{2\pi H}}.
\label{defs}
\end{equation}
The first Friedmann equation (\ref{frd1}) can be expressed also as the 
following relation among the Bekenstein entropy $S_B\equiv{{2\pi a}\over n}E$, 
the Bekenstein-Hawking entropy $S_{BH}\equiv(n-1){V\over{4Ga}}$, the Hubble 
entropy $S_H$ and $S_Q\equiv {{2\pi a}\over n}\cdot {1\over 2}\Phi Q$:
\begin{equation}
S^2_H=2(S_B-S_Q)S_{BH}-kS^2_{BH}.
\label{entrels}
\end{equation}
For the $k=1$ case, this can be written as the following quadratic relation:
\begin{equation}
S^2_H+(S_B-S_Q-S_{BH})^2=(S_B-S_Q)^2,
\label{quadrel}
\end{equation}
which generalizes the one in Ref. \cite{ver1}.  However, $S_B-S_Q$ does not 
remain constant during the cosmological evolution, as can be checked by 
applying Eq. (\ref{enrgcsv}) along with the equation of state $p=\rho/n$.  
So, one cannot draw a circular diagram with the constant radius $S_B-S_Q$ to 
study the time evolution of $S_H$ and $S_{BH}$ unlike the case in Ref. 
\cite{ver1}.  

We study thermodynamics of the CFT at the moment when the brane crosses the 
black hole horizon $a=a_H$, defined as the largest root of $h(a)=0$, i.e., 
\begin{equation}
{a^2_H\over L^2}+k-{{w_{n+1}M}\over a^{n-1}_H}+{{nw^2_{n+1}Q^2}
\over{8(n-1)a^{2n-2}_H}}=0.
\label{roothor}
\end{equation}
From Eqs. (\ref{frdeq1},\ref{roothor}), we see that 
\begin{equation}
H^2={1\over L^2}\ \ \ \ \ \ \ {\rm at}\ \ \ \ \ \ a=a_H.
\label{hblhor}
\end{equation}
The total entropy $S$ of the CFT remains constant, but the entropy density,
\begin{equation}
s\equiv{S\over V}=(n-1){a^n_H\over{4GLa^n}},
\label{entden}
\end{equation}
varies with time.  Making use of Eq. (\ref{hblhor}), we see that $s$ at 
$a=a_H$ can be expressed in terms of $H$ in the following form:
\begin{equation}
s=(n-1){H\over{4G}}\ \ \ \ \ \ \ {\rm at}\ \ \ \ \ \ a=a_H,
\label{entdenhor}
\end{equation}
which implies
\begin{equation}
S=S_H\ \ \ \ \ \ \ {\rm at}\ \ \ \ \ \ a=a_H.
\label{hentsat}
\end{equation}
Making use of the general formula $H^2=\sigma^2-h(a)/a^2$ that follows from 
Eqs. (\ref{costim},\ref{eqbranact}), we see that the CFT temperature $T=
h^{\prime}(a_H)L/(4\pi a_H)$ at $a=a_H$ can be expressed in terms of $H$ and 
$\dot{H}$ in the following way:
\begin{equation}
T=-{\dot{H}\over{2\pi H}}\ \ \ \ \ \ \ {\rm at}\ \ \ \ \ \ a=a_H.
\label{tmphor}
\end{equation}
Eq. (\ref{fredcft2}) along with Eqs. (\ref{hentsat},\ref{tmphor}) implies
\begin{equation}
E_C=kE_{BH}\ \ \ \ \ \ \ {\rm at}\ \ \ \ \ \ a=a_H,
\label{enrel}
\end{equation}
where $E_C$ is the Casimir energy defined as
\begin{equation}
E_C\equiv n(E+pV-\Phi_HQ-TS).
\label{casendef}
\end{equation}
So, even for the case of charged AdS black holes, for any values of $k$,  
thermodynamic quantities of the CFT take the forms simply expressed in terms 
of the Hubble parameter and its time derivative, when the brane crosses the 
black hole horizon.  

Thermodynamic quantities of the CFT satisfy the first law of thermodynamics,
\begin{equation}
TdS=dE-\Phi_HdQ+pdV,
\label{1stlaw}
\end{equation}
which can be re-expressed in terms of the densities as
\begin{equation}
Tds=d\rho-\Phi_Hd\rho_Q+n(\rho+p-\Phi_H\rho_Q-Ts){{da}\over a},
\label{1stlawden}
\end{equation}
making use of $dV=nVda/a$.  The combination $\rho+p-\Phi_H\rho_Q-Ts$ measures 
the subextensive contribution to the thermodynamic system.  To find the 
expression for the combination, we write the energy density of the CFT in 
the following way:
\begin{equation}
\rho={{na^n_H}\over{16\pi G_{n+2}a^{n+1}}}\left({a_H\over L}+k{L\over a_H}
+{{nw^2_{n+1}Q^2L}\over{8(n-1)a^{2n-1}_H}}\right),
\label{endenexp}
\end{equation}
and make use of the equation of state $p=\rho/n$, which holds for CFTs.  
The resulting expression is
\begin{equation}
{n\over 2}(\rho+p-\Phi_H\rho_Q-Ts)=k{\gamma\over a^2},
\label{combexp}
\end{equation}
where the Casimir quantity $\gamma$ is given by
\begin{equation}
\gamma={{n(n-1)a^{n-1}_H}\over{16\pi Ga^{n-1}}}.
\label{gamma}
\end{equation}
So, the Casimir energy (\ref{casendef}) of the CFT is
\begin{equation}
E_C={{kn(n-1)a^{n-1}_H{\rm Vol}(M^n)}\over{8\pi Ga}},
\label{caseng}
\end{equation}
which does not have explicit dependence on $Q$.  The entropy density 
(\ref{entden}) of the CFT can be expressed in terms of $\gamma$ and the 
densities as
\begin{equation}
s^2=\left({{4\pi}\over n}\right)^2\gamma\left(\rho-\textstyle{1\over 2}\Phi_H
\rho_Q-k{\gamma\over a^2}\right).
\label{entdenexp}
\end{equation}
By making use of Eq. (\ref{entdenhor}), we can show that the entropy density 
formula (\ref{entdenexp}) reproduces the first Friedmann equation (\ref{frd1}) 
when the brane crosses the black hole horizon.  Also, making use of Eqs. 
(\ref{entdenhor},\ref{tmphor}), we can show that Eq. (\ref{combexp}) 
reproduces the second Friedmann equation (\ref{frd2}) when $a=a_H$.  We have 
thus shown that the results of Ref. \cite{ver2} can be extended to the 
charged topological AdS black hole cases.  

From the entropy density formula (\ref{entdenexp}), we obtain the following 
generalized Cardy-Verlinde formula for the CFT with the R-charge $Q$:
\begin{equation}
S=\sqrt{{{2\pi a}\over n}S_C[2(E-\textstyle{1\over 2}\Phi_HQ)-E_C]},
\label{newentfrml}
\end{equation}
where the Casimir entropy $S_C$ and the Casimir energy $E_C$ are defined as
\begin{eqnarray}
S_C&\equiv&(n-1){{a^{n-1}_H{\rm Vol}(M^n)}\over{4G}},
\cr
E_C&\equiv&{{kn}\over{2\pi a}}S_C=kn(n-1){{a^{n-1}_H{\rm Vol}(M^n)}
\over{8\pi Ga}}.
\label{defscec}
\end{eqnarray}
Note, the Casimir energy $E_C$ in the above generalized Cardy-Verlinde 
formula is positive, zero and negative for $k=1,0,-1$, respectively, 
whereas the Casimir entropy $S_C$ is always positive for any $k$.  The 
Casimir entropy depends implicitly on $M$, $Q$ and $k$ through $a_H$ 
satisfying Eq. (\ref{roothor}).  The generalized Cardy-Verlinde formula 
(\ref{newentfrml}) coincides with the cosmological Cardy formula 
(\ref{fredcft1}) when the brane crosses the black hole horizon $a=a_H$.  
The generalized Cardy-Verlinde formula (\ref{newentfrml}) can be written 
also as the following relation:
\begin{equation}
S^2=2(S_B-S_{Q_H})S_C-kS^2_C,
\label{modcvrel}
\end{equation}
where $S_{Q_H}\equiv {{2\pi a}\over n}\cdot{1\over 2}\Phi_HQ$.  This relation 
and the relation (\ref{entrels}) among the cosmological entropy bounds 
coincide when $a=a_H$.  

We now discuss the cosmological holographic bounds generalized to the case 
of the charged topological AdS black holes.  The above expressions for the 
generalized Cardy-Verlinde formula suggest that the criterion for 
distinguishing between a weakly and a strongly self-gravitating universes 
should be modified to
\begin{eqnarray}
E-{1\over 2}\Phi_HQ\leq E_{BH} \ \ \ \Leftrightarrow\ \ \ S_B-S_{Q_H}
\leq S_{BH} \ \ \ \ \ \ \ &{\rm for}&\ \ \ \ \ \ Ha\leq\sqrt{2-k}
\cr
E-{1\over 2}\Phi_HQ\geq E_{BH} \ \ \ \Leftrightarrow\ \ \ S_B-S_{Q_H}\geq 
S_{BH} \ \ \ \ \ \ \ &{\rm for}&\ \ \ \ \ \ Ha\geq\sqrt{2-k}.
\label{wkstrgcrt}
\end{eqnarray}
We further propose that the cosmological bound on $S_C$ conjectured in Ref. 
\cite{ver1} continues to hold without modification, even when $Q\neq 0$:
\begin{equation}
S_C\leq S_{BH},
\label{cnjctcb}
\end{equation}
which can be inferred from the fact that the expression for the Casimir 
entropy in Eq. (\ref{defscec}) does not explicitly depend on $Q$, as well 
as on $k$.  From the explicit expression (\ref{defscec}) for $S_C$ and the 
definition for $S_{BH}$, we see that the cosmological bound (\ref{cnjctcb}) 
implies $a_H\leq a$.  So, the cosmological bound (\ref{cnjctcb}) is 
saturated when $a=a_H$, i.e., when the brane crosses the black hole horizon, 
at which moment the entropy $S$, the temperature $T$ and the Casimir energy 
$E_C$ of the CFT are given respectively by Eqs. 
(\ref{hentsat},\ref{tmphor},\ref{enrel}).  
We argue that the conjectured cosmological bound (\ref{cnjctcb}) implies 
the Hubble entropy bound and the modified Bekenstein bound for a strongly 
and a weakly self-gravitating cases, respectively.  For the strongly 
self-gravitating case, we see from Eqs. (\ref{wkstrgcrt},\ref{cnjctcb}) that 
$S_C\leq S_{BH}\leq S_B-S_{Q_H}$.  From Eq. (\ref{modcvrel}) we see that $S$ 
is a monotonically increasing function of $S_C$ in the interval $S_C\leq S_B
-S_{Q_H}$ for any values of $k$, as long as $S_B>S_{Q_H}$.  Thus, $S$ reaches 
its maximum value when $S_C=S_B-S_{Q_H}$ and therefore $S_C=S_{BH}$, for 
which $S=S_H$ (as can be seen from Eqs. (\ref{entrels},\ref{modcvrel}) with 
$S_C=S_{BH}$ and $a=a_H$).   The conjectured cosmological bound 
(\ref{cnjctcb}) therefore implies the Hubble entropy bound even for the 
$Q\neq 0$ case:
\begin{equation}
S\leq S_H  \ \ \ \ \ \ \ \ \ \ {\rm for}\ \ \ \ \ \ \ \ \ Ha\geq\sqrt{2-k}.
\label{entbnd}
\end{equation}
For the weakly self-gravitating case, we have $S_C\leq S_B-S_{Q_H}\leq S_{BH}$ 
for $k=1$, if it is assumed that $E_C\leq E-{1\over 2}\Phi_HQ$.  So, we have 
the following modified Bekenstein bound:
\begin{equation}
S\leq S_B-S_{Q_H} \ \ \ \ \ \ \ \ \ \ {\rm for}\ \ \ \ \ \ \ \ \ 
Ha\leq 1,
\label{modbekbnd}
\end{equation}
which is saturated when $S_C=S_B-S_{Q_H}$.  For any values of $k$, if we 
assume that $S_C\leq S_B-S_{Q_H}$ (which can be interpreted as the  
holographic upper limit on the degrees of freedom of the dual CFT, generalized 
to the case with an R-charge), then we also have $S_C\leq S_B-S_{Q_H}\leq 
S_{BH}$ for any $k$.  The Bekenstein bound is then generalized to
\begin{equation}
S\leq\sqrt{2-k}(S_B-S_{Q_H}) \ \ \ \ \ \ \ \ \ \ {\rm for}\ \ \ \ \ \ \ \ \ 
Ha\leq\sqrt{2-k},
\label{genbekbnd}
\end{equation}
which is saturated when $S_C=S_B-S_{Q_H}$.

\end{document}